\documentstyle{elsart}

\begin{document}
{\small \hfill  INP~MSU-99-45/603, TTP99-52}
 
\begin{frontmatter}
\title{The criterion of irreducibility of multi--loop Feynman integrals.
}


\author[A]{P.\,A.\,BAIKOV \thanksref{AA}}

\thanks[AA]{Supported in part by INTAS (grant YSF-98-173),
Volkswagen Foundation (grant No.~I/73611)
and RFBR (grant 98-02-16981);
e-mail: baikov@theory.npi.msu.su}

\address[A]{Institute of Nuclear Physics, Moscow State University,
Moscow~119899, Russia}

\begin{abstract}
The integration by parts recurrence relations allow to reduce
some Feynman integrals to more simple ones (with some lines
missing). Nevertheless the possibility of such reduction for the given
particular integral was unclear. The recently proposed technique for
studying the recurrence relations as by-product provides with simple
criterion of the irreducibility.
\end{abstract}

\end{frontmatter}

\section{Introduction.}

The growing accuracy of the experimental tests of the Standard Model of
particle physics
makes neseccary  calculation of higher order quantum  corrections.
The  latter, in turn,  are expressed through  so called multi-loop
Feynman integrals.

For example, the calculation of the $O(\alpha_s^2)$ correction to
$R(s)$ (cross-section $e^+e^- \rightarrow (hadrons)$)
demands calculation of the massive 3--loop propagator diagrams.
At present such diagrams cannot be evaluated explicitly, but it is
possible to construct the reliable approximation, considering their
expansions
in various kinematical regions \cite{approx}. After some algebraic
manipulations
the coefficient in these expansions can be related to a huge number 
(of oder of millions) of 3-loop propagator massless and vacuum massive
diagrams, which differ from each other by degrees of their denominators.

Fortunately, one need not calculate all these integrals separately,
because it is possible (using the integration by part
algorithm \cite{ch-tk}) find algebraical relations between these
integrals. Indeed, suppose we need to evaluate the L-loop Feynman integral
in the dimensional regularization:
\begin{equation}
B(\underline{n})\equiv
\int \cdots \int \frac{d^Dp_1\ldots d^Dp_L} 
{D_1^{n_1}\ldots D_N^{n_N}},
\label{eqbn}
\end{equation}
where $p_i$ (${i=1,\ldots,L}$) are loop momenta 
and $D_a=
A^{ij}_a p_i\cdot p_j - m_a^2$ (a=1,\ldots,N).
According to general rules of d-dimensional integration we can integrate
by part in this integral or, in other words, the integral of total
derivative according to loop momenta should vanish. From other side we can
evaluate this derivative as linear combination of the $B(\underline{n})$
with $n_i$ shifted by $\pm1$:
\begin{equation}
0=\int \cdots \int (\partial/\partial p_i)\cdot p_k
\frac{d^Dp_1\ldots d^Dp_L} 
{D_1^{n_1}\ldots D_N^{n_N}}=R(I^+,I^-)B(\underline{n}),
\label{eqrr}
\end{equation}
where  ${\bf I}^-_c B(.., n_c,..)\equiv B(..,n_c-1,..)$
and ${\bf I}^+_c B(.., n_c,.. )\equiv n_c B(.., n_c+1,..)$.

One can use these relations to relate an integral with some values of
$n_i$ to more simple integrals.
In particular, the millions of integrals appeared in the $O(\alpha_s^2)$
calculations mentioned above can be related with very
few integrals (6 massless propagator and 3 massive vacuum integrals),
which should be calculated explicitly.

The most important step in this method is to find the proper combination
of the relations of (\ref{eqrr}) type, which allow to reduce the given
integral
to the desirable set of integrals. In other words, one should represent
the relations (\ref{eqrr}) in recursive form, which allows to reduce each
index
$n_i$ to basic value (usually 0 or 1). This problem was solved for some
important cases \cite{alg}, but there is no systematic way for the
general case. It is even unknown, is it possible to reduce the given
integral to more simple integrals.

The answer to the last question is the main subject of the
present paper. In the next section the general form of the irreducibility
criterion is suggested. Then its application is illustrated by a 3-loop
example. In the last section some general remarks are presented.

\section{Irreducibility criterion.}

Suppose we want to check the possibility to represent, using the relations
(\ref{eqrr}), the given integral $B(\underline{n}_0)$ as linear
combination of integrals from the set $\{ B(\underline{n}_i),
i\in(1,...,k)\}$:
\begin{equation}
0=B(\underline{n}_0) - \sum_{i=1}^k f_i B(\underline{n}_i).
\label{eqstar}
\end{equation}
The relations (\ref{eqstar}) (if exist) should be the sequence of the
relations (\ref{eqrr}) (in fact, their linear combinations).

Let us assume that we found the other solution of (\ref{eqrr}) in form of
some function $s(\underline{n})$. Any such function should also fit the
(\ref{eqstar})
as the sequence of the relations (\ref{eqrr}).
Suppose we are able to construct some specific $s(\underline{n})$ with the
properties
\begin{equation}
s(\underline{n}_0)\neq0, \quad 
s(\underline{n}_i)=0\ \mbox{for}\ i\in(1,...,k).
\label{eqnorm}
\end{equation}
Such $s(\underline{n})$ evidently cannot fit (\ref{eqstar}) and hence
(\ref{eqstar})
cannot be the sequence of (\ref{eqrr}). So we got the sufficient criterion
of irreducibility: the $B(\underline{n}_0)$ cannot be represented as
linear combination of integrals from the set \{$B(\underline{n}_i)$,
$i\in(1,...,k)$\} if exist the partial solution of (\ref{eqrr}) with
properties (\ref{eqnorm}).

In practice, one can construct such solutions using the method of finding
the explicit solutions of the recurrence relations for Feynman integrals,
proposed in \cite{me}.
The key idea is to represent these solution in the form of auxlary
integral
\begin{equation}
s(\underline{n})=\int \frac{dx_1...dx_N}{x_1^{n_1}...
x_N^{n_N}}g(x_i).
\label{eqsol}
\end{equation}
One can check that action of the $R(I^+,I^-)$ on (\ref{eqsol}) leads to 
\begin{equation}
R(I^-,I^+) s(\underline{n})=\int
\frac{dx_1...dx_N}{x_1^{n_1}... x_N^{n_N}} R(x_i,\partial /
\partial x_i) g(x_i) + (\mbox{surface terms}).
\label{eqrrsol}
\end{equation}
So, if one choose the integrand as solution of the
\begin{equation}
R(\partial_i,x_i)g(x_i)=0
\label{eqrrg}
\end{equation}
and cancel the surface terms by proper choose of integration contour,
one can fit $R(I^+,I^-)s(\underline{n})=0$.

As it shown in \cite{me}, the (\ref{eqrrg}) can be solved for the general
case of multi-loop Feynman integral with arbitrary number of legs and with
arbitrary masses,
and the corresponding $g(x_i)$ can be represented as product of two
polinoms in $x_i$, each polinom raised to non-integer degree (see \cite{me}
for details).

\section{Example.}

As example, let us consider the master 3-loop massless
non-planar integral. 
This integral is supposed to be irreducible because of practical
failure to simplify it. To the best of author's knowledge, no 
proof has been  found.

So, let us check the possibility to reduce this integral to the linear
combination of the simpler integrals (with at least one line missing):

\setlength{\unitlength}{0.05mm}
\begin{picture}(500,200)
\put(50,100){\line( -1, 0){30}}
\put(50,100){\line( 1, 1){100}}
\put(50,100){\line( 1,-1){100}}
\put(150,200){\line( 1, -1){90}}
\put(350,0){\line( -1, 1){90}}
\put(150,200){\line( 1,  0){200}}
\put(150,  0){\line( 1,  1){200}}
\put(150,  0){\line( 1,  0){200}}
\put(450,100){\line(-1,-1){100}}
\put(450,100){\line(-1, 1){100}}
\put(450,100){\line( 1, 0){30}}
\put(230,220){\small 2}
\put(230,-70){\small 5}
\put(50,150){\small 1}
\put(50, 0){\small 6}
\put(410,150){\small 3}
\put(410, 0){\small 4}
\put(155, 90){\small 7}
\put(305, 90){\small 8}
\end{picture}
\begin{picture}(220,200)
\put(0,75){$=c_1(d)$}
\end{picture}
\begin{picture}(500,200)
\put(50,100){\line( -1, 0){30}}
\put(50,100){\line( 1, 1){100}}
\put(50,100){\line( 1,-1){100}}

\put(50,100){\line( 1, 0){190}}
\put(350,0){\line( -1, 1){90}}

\put(150,200){\line( 1,  0){200}}
\put(150,  0){\line( 1,  1){200}}
\put(150,  0){\line( 1,  0){200}}

\put(450,100){\line(-1,-1){100}}
\put(450,100){\line(-1, 1){100}}
\put(450,100){\line( 1, 0){30}}
\end{picture}
\begin{picture}(220,200)
\put(0,75){$+c_2(d)$}
\end{picture}
\begin{picture}(500,200)
\put(50,100){\line( -1, 0){30}}
\put(50,100){\line( 1, 1){100}}
\put(50,100){\line( 1,-1){100}}

\put(150,200){\line( 1, -1){200}}
\put(150,200){\line( 1,  0){200}}
\put(150,  0){\line( 0,  1){200}}
\put(150,  0){\line( 1,  0){200}}

\put(450,100){\line(-1,-1){100}}
\put(450,100){\line(-1, 1){100}}
\put(450,100){\line( 1, 0){30}}
\end{picture}
\begin{picture}(250,200)
\put(0,75){$+\ldots$}
\end{picture}

where numbers numerate the lines. Let us define denominators
as
$$
D_1=(l+p+q)^2,\ 
D_2=(l+k+p+q)^2,\ 
D_3=(k+p+q)^2,\ 
D_4=(k+p)^2,\ 
$$
$$
D_5=p^2,\ 
D_6=(l+p)^2,\ 
D_7=k^2,\ 
D_8=l^2,\ 
D_9=2\ l\cdot k.
$$
The scalar products can be expressed through denominators in the following
way:
$$
k^2=D_7,\ 
k\cdot l=\frac{1}{2}D_9,\ 
k\cdot p=\frac{1}{2}(D_4 - D_5 - D_7),\ 
l^2=D_8,\ 
p^2=D_5,
$$
$$
k\cdot q=\frac{1}{2}(D_2 - D_1 - D_4 + D_5 - D_9),\ 
l\cdot q=\frac{1}{2}(D_2 - D_3 + D_5 - D_6 - D_9),
$$
\begin{equation}
l\cdot p=\frac{1}{2}(D_6 - D_5 - D_8),\ 
p\cdot q=\frac{1}{2}(D_1 - D_2 + D_3 - D_5 + D_9-q^2).
\label{eqpp}
\end{equation}
According to \cite{me}, for this particular massless propagator-type case
the function $g(\underline{x})$ will read
\begin{equation}
g(x_i)=(q^2)^{1-d/2}\det_{kl}\Big((p_k\cdot p_l)(x_i)\Big)^{d/2-5/2},
\label{eqg}
\end{equation}
where $p_k=(p,k,l,q)$ and $(p_k\cdot p_l)(x_i)$ means (\ref{eqpp})
with substitution $D_i=x_i$.

Let us construct $s(\underline{n})$ with property $s(\underline{n})=0$
if some of $(n_1,...,n_8)$ less or equal to 0. The natural way to fit this
condition is to chose the integration contours for $(x_1,...,x_8)$ as
small circles around zero. In this case, according to Cauchi theorem,
the integrations will lead to calculation of the $(n_i-1)$ coefficient in
the Taylor expansion of the integrand (we omit for simplicity the overall
factor):
\begin{equation}
s(\underline{n})\propto
\int \frac{dx_9}{x_9^{n_9}}
\left[\frac{\partial_1^{n_1-1}\dots\partial_8^{n_8-1}}
{(n_1-1)!\dots(n_8-1)!}
P(x_i)^{d/2-5/2}
\right]
\Bigg|_{x_1,\dots,x_8=0}.
\label{eqs9}
\end{equation}
Note that according to (\ref{eqpp},\ref{eqg})  
$P(x_i)\big|_{x_1,\dots,x_8=0}\propto x_9^2(q^2-x_9)^2$. It means that
remaining $x_9$ integral can be expressed through Pochhammer's
symbols $(a)_n\equiv\Gamma(a+n)/\Gamma(a)$:
\begin{equation}
\int dx_9 x_9^k \big(x_9^2(q^2-x_9)^2\big)^{d/2-5/2}\propto
(d-4)_k/(2d-8)_k.
\label{eqs10}
\end{equation}
Formulas (\ref{eqs9},\ref{eqs10}) define $s(\underline{n})$ with
desired properties (in particular, $s(1,..,1,0)$ $=1$).
It means, according to general statement, that integration by
part relations can not reduce master 3-loop massless non-planar integral
$B(1,..,1,0)$ to linear combination of more simple integrals.

\section{Final remarks and Conclusion.}

In this paper the sufficient criterion of the irreducibility was
suggested:
the given integral is irreducible if the specific solution of the recurrence 
relations (with properties (\ref{eqnorm})) exists. Formally one can prove
the
reverse statement with the following restriction: suppose the given
integral $B(\underline{n}_0)$ is irreducible and we are able to construct
the procedure which reduce any integral to the given integral and some
others:
\begin{equation}
B(\underline{n})=
c_0(\underline{n}) B(\underline{n}_0)+
c_1(\underline{n}) B(\underline{n}_1)+..
\label{eqcc}
\end{equation}
Then the coefficient $c_0(\underline{n})$ near the given integral will
serve as the separating function for this integral. 
Proof: acting by recurrence relations on both sides of (\ref{eqcc}) we got
zero on the left side, and linear combination of basic integrals on the
right side. The coefficient near the given integral
$R_{ik}c_0(\underline{n})$ should be zero because otherwise this integral
will be linear combination of others. It means that $c(\underline{n})$ is
the solution of (\ref{eqrr}). Its "separating" initial conditions are
fitted by construction. 

So formally the existence of a separating solution is equivalent to
irreducibility. But from practical point of view to prove irreducibility
one should explicitly construct this solution. Our experience (up to
4-loop) shows that representation (\ref{eqsol}) provides with such
solutions, although at 4-loop level it demands some efforts to find the
proper integration contours.

The author is grateful to J.H.Kuhn, K.G.Chetyrkin for very useful
discussions and for the hospitality in the physics department of the
Karlsruhe University where essential part of this work was done.


\begin{thebibliography}{99}
\bibitem{approx} %
P.A. Baikov and D.J. Broadhurst,
  {\it Proceedings of the 4th International Workshop on Software
  Engineering and
  Artificial Intelligence for High Energy and Nuclear Physics (AIHENP95)},
  Pisa, Italy, 3-8 April 1995, 167, hep-ph/9504398; \\
K.G.~Chetyrkin, J.H.~Kuhn and M.~Steinhauser,
Nucl. Phys. {\bf B482}, (1996), 213.
\bibitem{ch-tk} %
K.\,G.\,Chetyrkin and F.\,V.\,Tkachov, Nucl.\,Phys.\,{\bf B192}
(1981) 159;\\ 
F.\,V.\,Tkachov, Phys.\,Lett.\,{\bf B100} (1981) 65.
\bibitem{alg} %
S.\,A.\,Larin, F.\,V.\,Tkachev and J.\,A.\,M.\,Vermaseren, 
NIKHEF--H/91--18 (1991);\\
D.\,J.\,Broadhurst, Z.\,Phys.\,{\bf C54} (1992) 599;\\
L.\,V.\,Avdeev, Comput.\,Phys.\,Commun. {\bf 98} (1996) 15.\\
O.\,V.\,Tarasov, Nucl.\,Phys.\,{\bf B502} (1997) 455;
\bibitem{me} %
P.\,A.\,Baikov, Phys.\,Lett.\,{\bf B385}, (1996) 404;
Nucl.~Instr.~\& Methods~{\bf A389} (1997) 347;\\
P.\,A.\,Baikov, M. Steinhauser, Comput.\,Phys.\,Commun. {\bf
115} (1998) 161.
\end{thebibliography}
\end{document}